\documentclass{article}
\usepackage{graphicx}

\begin{document}

\title{Network Algorithmics and the Emergence of the Cortical Synaptic-Weight
Distribution}

\author{Andre~Nathan\\
Valmir~C.~Barbosa\thanks{Corresponding author (valmir@cos.ufrj.br).}\\
\\
Universidade Federal do Rio de Janeiro\\
Programa de Engenharia de Sistemas e Computa\c c\~ao, COPPE\\
Caixa Postal 68511\\
21941-972 Rio de Janeiro - RJ, Brazil}

\date{}

\maketitle

\begin{abstract}
When a neuron fires and the resulting action potential travels down its axon
toward other neurons' dendrites, the effect on each of those neurons is mediated
by the weight of the synapse that separates it from the firing neuron. This
weight, in turn, is affected by the postsynaptic neuron's response through a
mechanism that is thought to underlie important processes such as learning and
memory. Although of difficult quantification, cortical synaptic weights have
been found to obey a long-tailed unimodal distribution peaking near the lowest
values, thus confirming some of the predictive models built previously. These
models are all causally local, in the sense that they refer to the situation in
which a number of neurons all fire directly at the same postsynaptic neuron.
Consequently, they necessarily embody assumptions regarding the generation of
action potentials by the presynaptic neurons that have little biological
interpretability. In this letter we introduce a network model of large groups of
interconnected neurons and demonstrate, making none of the assumptions that
characterize the causally local models, that its long-term behavior gives rise
to a distribution of synaptic weights with the same properties that were
experimentally observed. In our model the action potentials that create a
neuron's input are, ultimately, the product of network-wide causal chains
relating what happens at a neuron to the firings of others. Our model is then of
a causally global nature and predicates the emergence of the synaptic-weight
distribution on network structure and function. As such, it has the potential to
become instrumental also in the study of other emergent cortical phenomena.

\end{abstract}

\newpage
The weight of a synapse between a neuron's axon and another's dendrite is
generally understood to be some measure of how influential an action potential
fired by the presynaptic neuron can be on the buildup of a such a potential in
the postsynaptic neuron. While the physical entities whose measurement can be
said to relate to synaptic weights are
various \cite{song00,vanrossum00,kepecs02,song05}, recent experimental work
involving measurements of the excitatory postsynaptic potential amplitude has
revealed that synaptic weights follow a long-tailed distribution that is
unimodal and peaks near the lowest voltage values \cite{song05}. Understanding
the processes that give rise to a distribution with these properties can be
greatly enhanced by the construction of mathematical models that take into
account the nature of each neuron involved (excitatory or inhibitory), the
nature of a synapse's plasticity in terms of how its weight changes in response
to inter-neuron signaling, and also the distribution of firings in time.
Predictive models have been built with varying degrees of
success \cite{song00,vanrossum00,rubin01}, the most successful ones drawing on
relatively well established knowledge regarding the proportion of inhibitory
neurons to be used and the rule to change synaptic weights \cite{vanrossum00}.

Invariably, though, these models have relied on examining one single
postsynaptic neuron toward which firing patterns are directed that in essence
seek to summarize the entire input history of the postsynaptic neuron by a
simple stochastic process. Arguably this history is one of the most important
elements in giving rise to the synaptic-weight distribution in a way that can be
understood biologically \cite{barbour07}, but in all current models there is no
choice but to summarize it beyond retrieval. This happens because the models are
all strictly local, allowing for no causal dependency between what happens at
two neurons unless they are no farther apart from each other than one single
synapse. The model we now introduce addresses this severe shortcoming by
combining a network structure and algorithm with the proven mathematical
elements of the previous models.

The new model has a structural component and an algorithmic one. The structural
component is a directed graph $D$ whose nodes correspond to neurons that can be
either excitatory or inhibitory. For $i$ and $j$ two distinct nodes such that at
least one of them is excitatory, an edge directed from $i$ to $j$ represents a
synapse with associated weight $w_{ij}$. No edge exists between two inhibitory
nodes \cite{abeles91}. The algorithmic component turns each node in $D$ into a
simple simulator of the corresponding neuron, employing message passing on the
edges along their directions to simulate the signaling through the corresponding
synapses when nodes fire. Collectively, the nodes behave as an asynchronous
distributed algorithm \cite{barbosa96}, here referred to as A, each executing a
simple procedure P whenever receiving a message, possibly sending messages
itself while executing P but remaining idle at all other times. Because nodes
only do any processing in this reactive manner, at least one node is needed that
initially executes P once without any incoming message to respond to and then
starts behaving reactively like the others. We call such a node an initiator.

At node $j$, let $v_j$ stand for the node's potential. Let also $v^0$ and
$v^\mathrm{t}$ be a node's rest potential and threshold potential, respectively,
the same for all nodes. The effect of running P is for $j$ to probabilistically
decide whether to fire and, if it does fire, to send messages on all outgoing
edges while setting $v_j$ to $v^0$. If P is run as the initial processing by an
initiator, then the firing occurs with probability $1$ and P involves no actions
other than the ones just described. If not, then let $i$ be the sender of the
triggering message. The firing occurs with probability
$\min\{1,(v_j-v^0)/(v^\mathrm{t}-v^0)\}$ after $v_j$ has been updated to either
$v_j+w_{ij}$ (if $i$ is excitatory) or $v_j-w_{ij}$ (if $i$ is inhibitory). Then
the weight $w_{ij}$ is considered for an update.

The updating of $w_{ij}$ seeks to mimic the commonly accepted generalization of
the Hebbian rule embodied in the spike-timing-dependent plasticity
principles \cite{abbott00,song00}, according to which the change incurred by a
synapse's weight depends on the extent to which there is a causal dependency of
what happens at a neuron upon the other's firing. As a general rule, the
synaptic weight is increased (potentiated) if the postsynaptic neuron fires in
response to the firing by the presynaptic neuron, decreased (depressed)
otherwise. In either case the amount of change to the synaptic weight depends on
how close in time the relevant firings are, becoming negligible with increasing
separation. Procedure P follows these principles by keeping track of the latest
firing by $j$ so that a decision can be made on whether to increase or decrease
$w_{ij}$. If $j$ does fire in response to the message received from $i$, then
$w_{ij}$ is increased. If it does not but the previous message received from any
source did cause $j$ to fire, then $w_{ij}$ is decreased. The weight $w_{ij}$
remains unchanged in all other cases. The actual amount of change to $w_{ij}$
depends on whether it is to be increased or decreased, and so does the nature of
the change (by a fixed amount or by proportion) \cite{bi98,bi01,kepecs02}. An
increase in $w_{ij}$ is implemented by setting $w_{ij}$ to
$\min\{1,w_{ij}+\delta\}$ with $\delta>0$, a decrease by setting $w_{ij}$ to
$(1-\alpha)w_{ij}$ with $0<\alpha<1$, thus ensuring that synaptic weights remain
in the $[0,1]$ interval if so started.

\begin{figure}
\centering
\scalebox{0.8}{\includegraphics{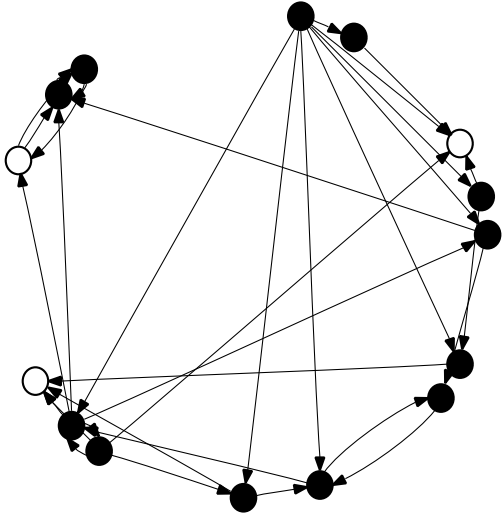}}
\caption{Network topology. Restricted to two dimensions for visual clarity, a
$D$ instance comprises nodes positioned randomly on a radius-$1$ circle and
edges, drawn as chords of the circle, that tend to be more abundant over lower
Euclidean distances. Excitatory nodes are represented by filled circles,
inhibitory nodes by empty circles.}
\label{fig:figure1}
\end{figure}

Running algorithm A starts with choosing one or more initiators, each of which
executes P and then starts behaving like all other nodes. At any time it may
happen that a node has more than one input message to process, in which case the
order in which they are taken is the order of message reception. Because this
order is in principle arbitrary, A is seen to acquire another degree of
indeterminacy, in addition to that which is already present owing to the
probabilistic decisions. We have conducted extensive computational
experimentation with A on a graph $D$ intended to model a simple cortex, in line
with significant recent work that draws on the theory of graphs to help solve
problems in
neuroscience \cite{sporns04,sporns05,achard06,bassett06,he07,honey07,
reijneveld07,sporns07,stam07}. We regard $D$ as a random graph but, unlike some
of the early work on cortical modeling by such graphs \cite{abeles91}, where
fully random graphs \cite{erdos59} were used, we let $D$ have a scale-free
structure \cite{newman05}, with parameter as suggested by some of the more
recent finds \cite{eguiluz05,vandenheuvel08}. Thus, a randomly chosen node $i$
in $D$ has $k$ outgoing edges with probability proportional to $k^{-1.8}$.
Moreover, inspired by recent work on the modeling of cortical
systems \cite{kaiser04a,kaiser04b}, we let each outgoing edge of $i$ lead to
another randomly chosen node $j$ with probability proportional to $e^{-2d}$,
where $d$ is the Euclidean distance between $i$ and $j$ when the nodes of $D$
are placed uniformly at random on a radius-$1$ sphere
(Figure~\ref{fig:figure1}), provided $i$ and $j$ are not both inhibitory.

All computational experiments have adhered to the methods described next, which
refer to sequences of $10\,000$ runs of algorithm A. The first run in a sequence
operates on initial node potentials and synaptic weights chosen randomly from
the intervals $[v^0,v^\mathrm{t}]$ and $[0,1]$, respectively, with $v^0=-15$ and
$v^\mathrm{t}=0$. Each subsequent run operates on the potentials and weights
left by the previous run. For $n$ the number of nodes in $D$, a new set of
$0.05n$ initiators is chosen randomly at the beginning of each run. A run of A
is implemented as a sequential program that selects the next node to be
processed randomly (first out of the group of initiators for their first
executions of P, then out of those nodes that have at least one message to be
received). A new run in a sequence is only started after the previous one has
died out (no more messages to be processed remain), which is guaranteed to
happen eventually with probability $1$. The remaining parameters used by
procedure P are $\delta=0.01$ and $\alpha=0.05$. All our results refer to
$50\,000$ independent sequences, of which each $500$ sequences correspond to a
new $D$ instance. A $D$ instance is constructed by first placing all nodes
uniformly at random on a radius-$1$ sphere, then selecting the number of
outgoing edges for each node. Nodes are then chosen to be excitatory or
inhibitory randomly, provided a certain proportion is respected, and the
destination of each edge is decided. The graph that is actually used in the run
sequences is the giant strongly connected component of $D$ \cite{dorogovtsev01},
so a directed path exists from any node to any other. For the connectivity
distribution and construction method in use this component comprises about
$0.95n$ nodes on average.

\begin{figure}
\centering
\scalebox{0.8}{\includegraphics{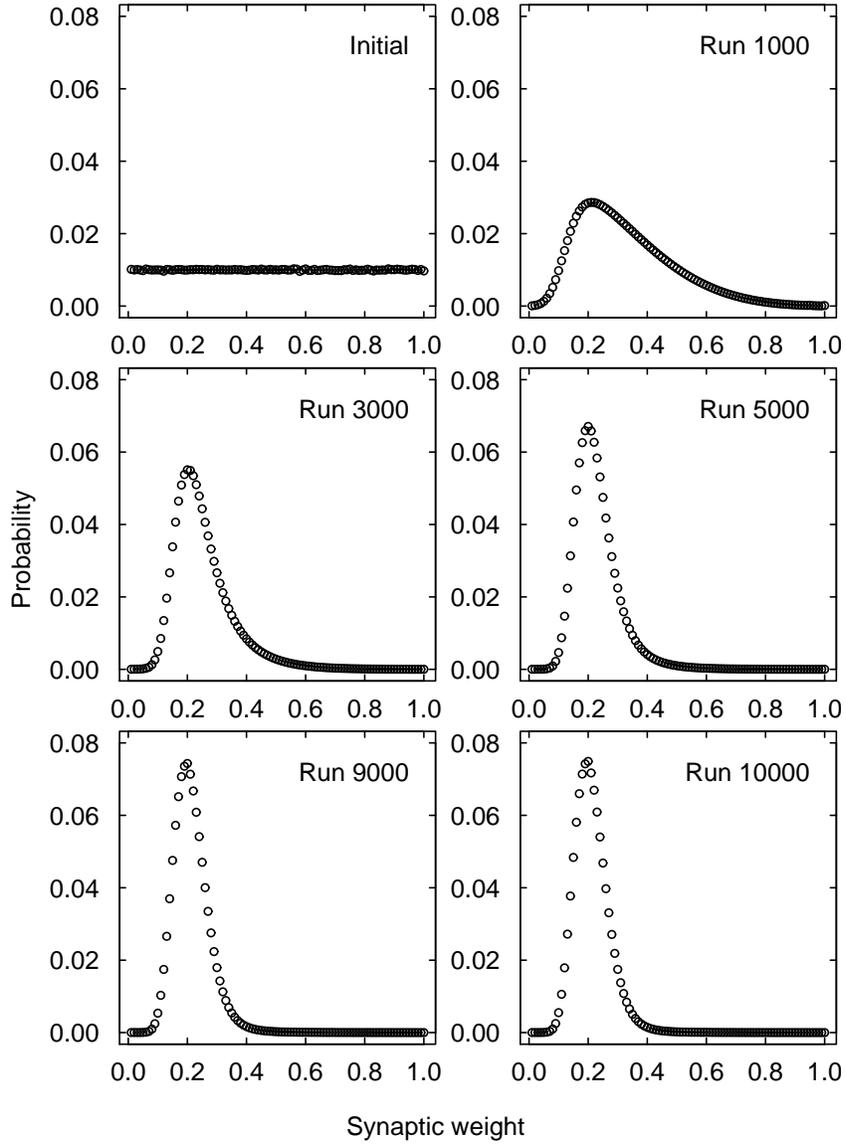}}
\caption{The synaptic-weight distribution, shown after selected runs of
algorithm A. Probabilities are binned to a fixed width of $0.01$.}
\label{fig:figure2}
\end{figure}

\begin{figure}
\centering
\scalebox{0.8}{\includegraphics{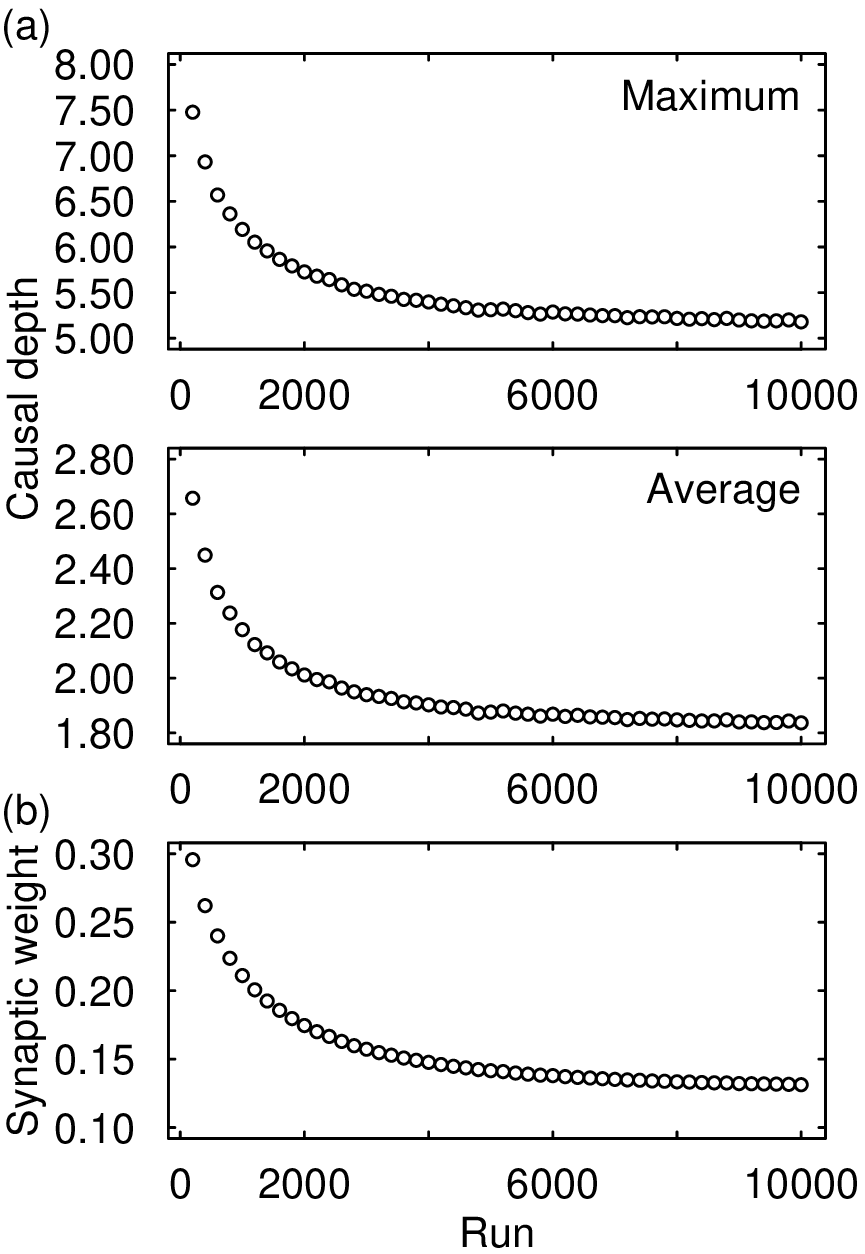}}
\caption{Causal depth of a message reception and associated synaptic weights.
The causal depth of a message reception is the size of its causal history, i.e.,
the number of firings that precede it along the chain of firings that begins at
some initiator when it fires for the first time, each preceding the next by
direct causation: given any two subsequent firings in this chain, the first
entails the sending of a message whose reception triggers the second. (a)
Maximum and average causal depth of terminal message receptions during the
course of each run. (b) Average weight (before updates) of the synapses involved
in the causal histories of terminal message receptions.}
\label{fig:figure3}
\end{figure}

Our results, here given for $n=1\,000$ and the well accepted proportion of
$0.2n$ inhibitory nodes \cite{abeles91,ananthanarayanan07}, show that the
synaptic-weight distribution becomes analogous to the distribution unveiled by
experimentation along the sequences of runs of algorithm A described above
(Figure~\ref{fig:figure2}). The process is gradual, leading the weights to
become relatively concentrated around a single low-value mode while still
allowing some residual probability to remain at the higher values. The long-term
distribution is seen to stabilize even as the weights continue to evolve, thus
suggesting the existence of an underlying weight dynamics whose effect on the
overall distribution is nevertheless practically imperceptible. The existence of
this persistent dynamics is revealed by the causal history of each terminal
message reception (one that does not lead to the firing of the receiver), which
can be significantly deep with respect to the relatively short average path of a
scale-free network \cite{newman01} [Figure~\ref{fig:figure3}(a)]. The sending of
every message by a non-initiator causes a synaptic weight to be increased,
unless it already equals $1$, but weight-$1$ synapses are very rare, especially
when arranged as a path in $D$. So the causal histories we have discovered do
indeed hint at the existence of a dynamics of weight evolution in which weights
both increase and decrease in complex patterns. Additional confirmation is
provided by the average weight of the synapses involved in the causal histories
of terminal message receptions, which is consistently less than $1$ and also
decreases throughout the runs as the synaptic-weight distribution settles
[Figure~\ref{fig:figure3}(b)].

\begin{figure}
\centering
\scalebox{0.8}{\includegraphics{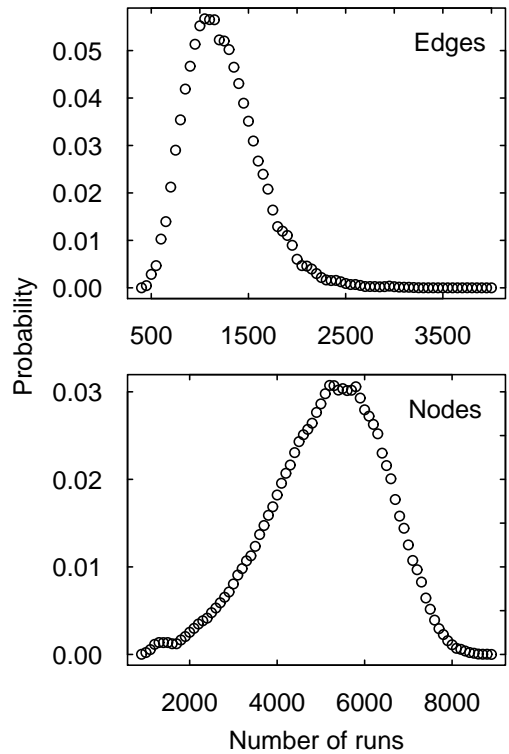}}
\caption{Final distributions of the number of runs in which an edge is traversed
or a node is reached. An edge is said to be traversed in a run when at least one
message is sent along it during the course of that run. A node is said to be
reached in a run when it receives at least one message during the course of that
run. Probabilities are binned to a fixed width of $50$ for edges, $100$ for
nodes.}
\label{fig:figure4}
\end{figure}

Every run in the sequences to which Figures~\ref{fig:figure2} and
\ref{fig:figure3} refer involves a new group of initiators and as such provides
new possibilities regarding the branching of causal histories and how they
affect firings and weight changes throughout the network. Monitoring the traffic
of messages as they traverse edges and reach nodes is then a means to do some
quantification of how the cascading runs, with their intermingling causal trees
rooted at many different initiators, cooperate in promoting the emergence of the
synaptic-weight distribution. We have found that the long-term distributions of
how many runs traverse an edge or reach a node (Figure~\ref{fig:figure4}),
allowing as they do for relatively high numbers with significant probabilities,
suggest that some sort of information integration is taking place among portions
of the network as the runs unfold. Perhaps such integration occurs in a sense
similar to that which has been theorized recently regarding the emergence of
higher functions such as consciousness \cite{balduzzi08}. If so, then network
algorithmics such as we have discussed may come to provide a powerful framework
to test the assumptions and eventual predictions of such theories.

\subsection*{Acknowledgments}

We acknowledge partial support from CNPq, CAPES, and a FAPERJ BBP grant.

\newpage
\bibliography{swd}

\begin{thebibliography}{10}

\bibitem{abbott00}
L.~F. Abbott and S.~B. Nelson.
\newblock Synaptic plasticity: taming the beast.
\newblock {\em Nature Neuroscience}, 3:1178--1183, 2000.

\bibitem{abeles91}
M.~Abeles.
\newblock {\em Corticonics: Neural Circuits of the Cerebral Cortex}.
\newblock Cambridge University Press, Cambridge, UK, 1991.

\bibitem{achard06}
S.~Achard, R.~Salvador, B.~Whitcher, J.~Suckling, and E.~Bullmore.
\newblock A resilient, low-frequency, small-world human brain functional
  network with highly connected association cortical hubs.
\newblock {\em Journal of Neuroscience}, 26:63--72, 2006.

\bibitem{ananthanarayanan07}
R.~Ananthanarayanan and D.~S. Modha.
\newblock Anatomy of a cortical simulator.
\newblock In {\em Proceedings of the 2007 ACM/IEEE Conference on
  Supercomputing}, page~3, 2007.

\bibitem{balduzzi08}
D.~Balduzzi and G.~Tononi.
\newblock Integrated information in discrete dynamical systems: motivation and
  theoretical framework.
\newblock {\em PLoS Computational Biology}, 4:e1000091, 2008.

\bibitem{barbosa96}
V.~C. Barbosa.
\newblock {\em An Introduction to Distributed Algorithms}.
\newblock The MIT Press, Cambridge, MA, 1996.

\bibitem{barbour07}
B.~Barbour, N.~Brunel, V.~Hakim, and J.~P. Nadal.
\newblock What can we learn from synaptic weight distributions?
\newblock {\em Trends in Neurosciences}, 30:622--629, 2007.

\bibitem{bassett06}
D.~S. Bassett and E.~Bullmore.
\newblock Small-world brain networks.
\newblock {\em Neuroscientist}, 12:512--523, 2006.

\bibitem{bi98}
G.~Q. Bi and M.~M. Poo.
\newblock Synaptic modifications in cultured hippocampal neurons: dependence on
  spike timing, synaptic strength, and postsynaptic cell type.
\newblock {\em Journal of Neuroscience}, 19:10464--10472, 1998.

\bibitem{bi01}
G.~Q. Bi and M.~M. Poo.
\newblock Synaptic modification by correlated activity: {H}ebb's postulate
  revisited.
\newblock {\em Annual Review of Neuroscience}, 24:139--166, 2001.

\bibitem{dorogovtsev01}
S.~N. Dorogovtsev, J.~F.~F. Mendes, and A.~N. Samukhin.
\newblock Giant strongly connected component of directed networks.
\newblock {\em Physical Review E}, 64:025101, 2001.

\bibitem{eguiluz05}
V.~M. Egu{\'{i}}luz, D.~R. Chialvo, G.~A. Cecchi, M.~Baliki, and A.~V.
  Apkarian.
\newblock Scale-free brain functional networks.
\newblock {\em Physical Review Letters}, 94:018102, 2005.

\bibitem{erdos59}
P.~Erd{\H{o}}s and A.~R{\'{e}}nyi.
\newblock On random graphs.
\newblock {\em Publicationes Mathematicae---Debrecen}, 6:290--297, 1959.

\bibitem{he07}
Y.~He, Z.~J. Chen, and A.~C. Evans.
\newblock Small-world anatomical networks in the human brain revealed by
  cortical thickness from {MRI}.
\newblock {\em Cerebral Cortex}, 17:2407--2419, 2007.

\bibitem{honey07}
C.~J. Honey, R.~K{\"{o}}tter, M.~Breakspear, and O.~Sporns.
\newblock Network structure of cerebral cortex shapes functional connectivity
  on multiple time scales.
\newblock {\em Proceedings of the National Academy of Sciences USA},
  104:10240--10245, 2007.

\bibitem{kaiser04b}
M.~Kaiser and C.~C. Hilgetag.
\newblock Modelling the development of cortical systems networks.
\newblock {\em Neurocomputing}, 58--60:297--302, 2004.

\bibitem{kaiser04a}
M.~Kaiser and C.~C. Hilgetag.
\newblock Spatial growth of real-world networks.
\newblock {\em Physical Review E}, 69:036103, 2004.

\bibitem{kepecs02}
A.~Kepecs and M.~C.~W. {van Rossum}.
\newblock Spike-timing-dependent plasticity: common themes and divergent
  vistas.
\newblock {\em Biological Cybernetics}, 87:446--458, 2002.

\bibitem{newman05}
M.~E.~J. Newman.
\newblock Power laws, {P}areto distributions and {Z}ipf's law.
\newblock {\em Contemporary Physics}, 46:323--351, 2005.

\bibitem{newman01}
M.~E.~J. Newman, S.~H. Strogatz, and D.~J. Watts.
\newblock Random graphs with arbitrary degree distributions and their
  applications.
\newblock {\em Physical Review E}, 64:026118, 2001.

\bibitem{reijneveld07}
J.~C. Reijneveld, S.~C. Ponten, H.~W. Berendse, and C.~J. Stam.
\newblock The application of graph theoretical analysis to complex networks in
  the brain.
\newblock {\em Clinical Neurophysiology}, 118:2317--2331, 2007.

\bibitem{rubin01}
J.~Rubin, D.~D. Lee, and H.~Sompolinsky.
\newblock Equilibrium properties of temporally asymmetric {H}ebbian plasticity.
\newblock {\em Physical Review Letters}, 86:364--367, 2001.

\bibitem{song00}
S.~Song, K.~D. Miller, and L.~F. Abbot.
\newblock Competitive {H}ebbian learning through spike-timing-dependent
  synaptic plasticity.
\newblock {\em Nature Neuroscience}, 3:919--926, 2000.

\bibitem{song05}
S.~Song, P.~J. Sj{\"{o}}str{\"{o}}m, M.~Reigl, S.~Nelson, and D.~B. Chklovskii.
\newblock Highly nonrandom features of synaptic connectivity in local cortical
  circuits.
\newblock {\em PLoS Biology}, 3:507--519, 2005.

\bibitem{sporns04}
O.~Sporns, D.~R. Chialvo, M.~Kaiser, and C.~C. Hilgetag.
\newblock Organization, development and function of complex brain networks.
\newblock {\em Trends in Cognitive Sciences}, 8:418--425, 2004.

\bibitem{sporns07}
O.~Sporns, C.~J. Honey, and R.~K{\"{o}}tter.
\newblock Identification and classification of hubs in brain networks.
\newblock {\em PLoS ONE}, 2:e1049, 2007.

\bibitem{sporns05}
O.~Sporns, G.~Tononi, and R.~K{\"{o}}tter.
\newblock The human connectome: a structural description of the human brain.
\newblock {\em PLoS Computational Biology}, 1:245--251, 2005.

\bibitem{stam07}
C.~J. Stam and J.~C. Reijneveld.
\newblock Graph theoretical analysis of complex networks in the brain.
\newblock {\em Nonlinear Biomedical Physics}, 1:3, 2007.

\bibitem{vandenheuvel08}
M.~P. {van den Heuvel}, C.~J. Stam, M.~Boersma, and H.~E. {Hulshoff Pol}.
\newblock Small-world and scale-free organization of voxel-based resting-state
  functional connectivity in the human brain.
\newblock {\em Neuroimage}, 43:528--539, 2008.

\bibitem{vanrossum00}
M.~C.~W. {van Rossum}, G.~Q. Bi, and G.~G. Turrigiano.
\newblock Stable {H}ebbian learning from spike timing-dependent plasticity.
\newblock {\em Journal of Neuroscience}, 20:8812--8821, 2000.

\end{thebibliography}
\bibliographystyle{plain}

\end{document}